\documentclass[12pt,a4paper]{amsart}
\usepackage[margin=3cm]{geometry}

\usepackage{graphicx}
\usepackage{psfrag} 
\usepackage{mathptmx}
\usepackage{amssymb}
\usepackage{amsmath}
\usepackage{color}
\usepackage[square,sort,comma,numbers]{natbib}

\usepackage[all]{xy}
\usepackage[mathscr]{euscript}
\usepackage{amscd}
\usepackage{epsfig}

\newtheorem{lemma}{Lemma}[section]
\newtheorem{theorem}[lemma]{Theorem}
\newtheorem{proposition}[lemma]{Proposition}
\newtheorem{corollary}[lemma]{Corollary}
\newtheorem{remark}[lemma]{Remark}
\newtheorem{definition}[lemma]{Definition}
\newtheorem{observation}[lemma]{Observation}

\newcommand{\pf}{\noindent{\em Proof: }}
\newcommand{\epf}{\hfill\hbox{\rule{3pt}{6pt}}\\}

\begin{document}
	
	\title{Tree-based unrooted phylogenetic networks}
	\author[Francis, Huber, Moulton]{A. Francis, K.T. Huber, and V. Moulton} 
	\address{Francis: Centre for Research in Mathematics, Western Sydney University, Australia.  e-mail: a.francis@westernsydney.edu.au}
	\address{Huber, Moulton: School of Computing Sciences, University of East Anglia, UK. e-mail: k.huber@uea.ac.uk, v.moulton@uea.ac.uk}

\begin{abstract}
	Phylogenetic networks are a generalization of phylogenetic trees that are used  
	to represent non-tree-like evolutionary histories that arise in organisms 
	such as plants and bacteria, or uncertainty in evolutionary histories. 
	An {\em unrooted} phylogenetic network
	on a nonempty, finite set $X$ of taxa, or {\em network}, is a  connected graph 
	in which every vertex has degree 1 or 3 and whose leaf-set is $X$.
	It is called a {\em phylogenetic tree} if the underlying graph is a tree.
	In this paper we consider properties of {\em tree-based networks}, that is,
	networks that can be constructed by adding edges into a phylogenetic tree.
	We show that although they have some properties in common with their 
	rooted analogues which have recently drawn much attention
	in the literature, they have some striking differences in terms of both their
	structural and computational properties. We expect that our results 
	could eventually have applications to, for example, detecting horizontal 
	gene transfer or hyrbridization which are important factors in the
	evolution of many organisms.
\end{abstract}
	
\maketitle

\noindent{\bf Keywords:} Phylogenetic tree, Phylogenetic network, Tree-based network, Hamiltonian path

\section{Introduction}

Let $X$ be a finite set with $|X| \ge 1$. 
An {\em unrooted phylogenetic network $N$ (on $X$)} 
(or {\em network $N$ (on $X$)} for short) is a connected graph $(V,E)$ with 
$X \subseteq V$, every vertex has degree 1 or 3, and the set of degree 1 
vertices (or {\em leaves}) is precisely $X$. A {\em phylogenetic tree} 
on $X$ is a network which is also a tree. Phylogenetic trees and 
networks are commonly used by biologists to represent the evolution of species; in this 
setting the set $X$ usually denotes a collection of species.
Networks have interesting mathematical and 
computational properties (see e.g. \cite{hmw16,g12,ssu}), and 
they can be generated from biological
data using software packages such as T-REX \cite{m01} and Splitstree \cite{hb06}.
In addition, networks have been used to study the genome fusion origin 
of eukaryotes \cite{r04} and as a tool in biogeography studies \cite{lm02}.

\begin{figure}[h]
	\begin{center}
		\includegraphics[width=7cm]{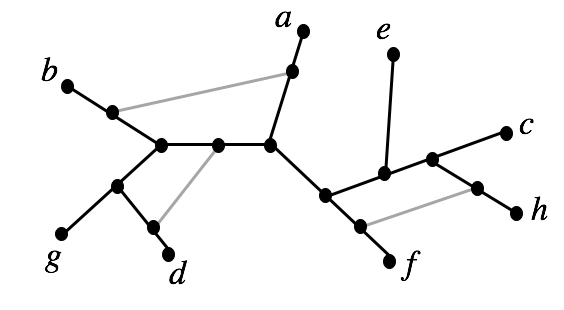}
	\end{center}
	\caption{A tree-based network that has been constructed from a
		phylogenetic tree with leaf-set $\{a,b,\dots,g\}$ by adding in 3 edges (in grey). Note that
		the tree is also a spanning tree for the network. }\label{addin}
\end{figure}

The T-REX software constructs networks (also
called reticulograms) by adding edges into a phylogenetic tree \cite{m01} 
(see e.g. Figure~\ref{addin}). Using this approach, many different networks can 
be constructed from starting with the collection of all phylogenetic trees. 
However, it is not possible to construct every possible network in this manner
(see for example Figure~\ref{f:level5} below). Indeed, the networks
that can be constructed in this way are of precisely the following type (cf. \cite{fss17}).

\begin{definition}
A network is {\em tree-based (on $X$)} if there is a spanning tree\footnote{Recall, a \emph{spanning tree} 
	of a graph $G$ is a connected subgraph of $G$ containing all vertices of $G$ and a minimal subset of the
	 edges} in $N$ whose leaf-set is equal to $X$. 
\end{definition}

Note that in the following 
we call any spanning tree in $N$ with leaf-set $X$ a {\em support-tree (for $N$)}.

Recently, a great deal of interest has been generated concerning 
rooted tree-based networks. These are leaf-labelled networks whose underlying graph is a 
directed acyclic graph with a single root which can be
constructed from a rooted phylogenetic tree by adding in extra arcs 
(see Section~\ref{build} for precise definitions). In particular, 
rooted tree-based networks were introduced in \cite{fs15} and 
their structural properties have been studied in \cite{fss17,h16,ji16,s16,z16}. 
In addition, various computational properties of these networks 
have been considered \cite{A16,fs15}.

Biologically, tree-based networks are a natural object of interest because 
their presentation of reticulations reflects assumptions about the underlying 
biological events (such as horizontal gene transfer or hybridization).  
Such events are regarded as occurring between taxonomic units (such as species) 
that also undergo vertical evolution; hence they are thought of as arcs between `tree' arcs.   
Biologists want to detect and understand horizontal evolution through events such as 
gene transfer and hybridization 
since this process plays an important role in the evolution of  many organisms (e.g.
in bacteria and plants).
Moreover, detecting horizontal evolution can be useful in applications
e.g. horizontal gene transfer is the primary mechanism for the spread of 
antibiotic resistance in bacteria \cite{G14}.

In this paper, we introduce tree-based networks in an unrooted setting, and 
present various results concerning them. 
Understanding such networks in an 
unrooted context has similar value to the study of phylogenetic trees in 
the unrooted context: we are able to allow for uncertainty in the placement of the root.  
As we shall see, although tree-based networks have certain
properties in common with their rooted analogues, they can behave quite differently
both in terms of their structural and computational properties. 

The outline of the paper is as follows. In the next section, we introduce
some relevant basic terminology and also present a way to decompose a
network into simpler, easier to understand pieces. In Section ~\ref{sec:recognize}, we
consider the computational problem of deciding whether or
not a given network $N$ is tree-based. For this we introduce two
novel decision problems and contrast our findings with the analogous
situation for rooted phylogenetic networks. In Section~\ref{sec:universal}, 
we show that
for every non-empty set $X$ there exists a tree-based network $N$ on
$X$ such that every phylogenetic tree $T$ on $X$ is a
base-tree for $N$, and we also establish an explicit relationship between rooted and unrooted tree-based networks (Theorem~\ref{rooted}).
In Section~\ref{sec:fully-tree-based}, we characterize level-1 networks
in terms of tree-based networks. We conclude with Section~\ref{sec:conclusion} where we
outline research directions that might be of interest.

\section{Decomposing tree-based networks}
\label{sec:decomp}

We begin by showing that networks can be decomposed into
simpler pieces, which can then be used to deduce properties
of the full network. Note that decomposition results have also 
been proven for rooted tree-based networks, although these are 
quite different in nature (see e.g. \cite{z16}).

We begin by presenting some definitions.
A \emph{cut-edge}, or \emph{bridge}, of a network is an edge 
whose removal disconnects the graph.  
A cut-edge is \emph{trivial} if one of the connected components induced by deleting the 
cut-edge is a vertex (which must necessarily be a leaf).
A \emph{simple} network is one all of whose cut-edges are trivial 
(so for instance, note that trees on more than 3 leaves are \emph{not} simple networks).
A \emph{blob} in a network is a maximal subgraph that has no cut-edge, and 
that is not a  vertex~\cite{g12}.  For example, the network in Figure~\ref{addin}
contains one non-trivial cut-edge and two blobs.

Now, given a network $N$ and a blob $B$ in $N$, we 
define a simple network $B_N$ by taking the union of $B$ and all 
cut-edges in $N$ incident with $B$ (the
leaf-set of $B_N$ is just the set of end vertices of these 
cut-edges that are not already a vertex in $B$).

\begin{proposition}\label{decompose}
	Suppose $N$ is a network. Then $N$ is tree-based if and only if $B_N$ is tree-based for every blob 
	$B$ in $N$.
\end{proposition}
\noindent{\em Proof:}
If $N$ is tree-based, then it contains a support-tree $T$. Since every cut-edge in $N$ must be contained in 
a support-tree, it follows that if $B$ is a blob in $N$ then $T$ must induce a spanning tree of $B_N$ 
that contains every vertex in $B_N$. Therefore $B_N$ is tree-based.

Conversely, if $B_N$ is tree-based for every blob $B$ 
in $N$, then by taking a support-tree in $B_N$ for
each blob $B$ in $N$, we can clearly construct a 
spanning tree for $N$ that contains all vertices in $N$. Therefore $N$ is tree-based.
\epf

Using the last result, we can immediately classify the tree-based networks
having a single leaf.

\begin{observation}\label{obs1}
	Suppose $N$ is a network on $\{x\}$. Then $N$ is tree-based if and only if $N=(\{x\},\emptyset)$.
\end{observation}

We now look in more detail at the cut-edges of a tree-based network.
A {\em split} of $X$ is a bipartition of $X$ into two non-empty sets. If we 
remove a cut-edge from a network, then in some
cases the two resulting graphs will induce a split of $X$. 
We now show that if $N$ 
is tree-based, then this is always the case.

\begin{lemma}\label{lem:cut-edge-split}
	If $N$ is a tree-based network, then every 
	cut-edge of $N$ induces a split of $X$. 
\end{lemma}
\noindent{\em Proof:} 
If we have a cut-edge of $N$ that does not induce a 
split of $X$, then it follows that
there must be some blob $B$ in $N$ such that $B_N$ is a network with one leaf. But then $B_N$ is not tree-based by 
Observation~\ref{obs1}. This is a contradiction by Proposition~\ref{decompose}.
\epf 

We call a network $N$ {\em proper} if every cut-edge induces 
a split of $X$. By Lemma~\ref{lem:cut-edge-split}, 
all tree-based networks are proper.

Interestingly, using Proposition~\ref{decompose}, we are able to 
now show that certain low complexity proper networks are always tree-based. We first make a useful observation.

\begin{lemma}\label{lem:andrew}
	Let $N$ be a network on $X$ with $|X|\ge 2$.  For any $x\in X$ let $N-x$
	denote the network obtained from $N$ by deleting $x$
	and its incident edge, and suppressing the 
	resulting degree 2 vertex.
	If $N-x$ is tree-based, then so is $N$.
\end{lemma}
\noindent {\em Proof:} 
Suppose $x\in X$ and $T$ is a support-tree for $N-x$. 
Let $v\in V(N)$ denote the vertex adjacent with $x$ that
was suppressed in the construction of $N-x$ and let $e\in E(N-x)$
denote the resulting edge in $N-x$. If $e\in E(T) $, 
then we can obtain a support-tree for
$N$ by subdividing $e$ with a new vertex $w$
and adding the edge $\{w,x\}$ to $T$.  If $e\not\in E(T)$, 
then, since $T$ is a support tree for $N-x$, $T$ 
must contain both vertices in $e$, say $v_1$ and $v_2$. 
Therefore, we can obtain 
a support tree for $N$ by adding  a new vertex $w$  and the
edges $\{v_1,w\}$ and $\{w,x\}$ (or indeed $\{v_2,w\}$ and $\{w,x\}$) to $T$. 
\epf

Suppose $N$ is a network on $X$ and $k\geq 0$ is an 
integer. Then $N$ is called a {\em level-k network} if at most $k$
edges have to be removed from each blob of $N$ to obtain a tree. 
For example, the network in Figure~\ref{addin} is a level-2 network. 

\begin{theorem}\label{level-4-tree-based}
	All proper level-4 networks are tree-based.
	Moreover, networks of level greater than 4 need not be tree-based.
\end{theorem}

\noindent {\em Proof:} 
First observe that by definition of level-$k$ network, the main claim applies to level-1, 2, 3, and 4 networks.

Note that since $N$ is a proper network 
on $X$ it must contain at least two leaves. Also note that 
the theorem is straight-forward to check in case $N$ is level-0 or level-1. 

In case $N$ has level $2 \le k \le 4$, since $N$ is proper, by Proposition~\ref{decompose}
it suffices to prove that every simple, level-4 network with two leaves is tree-based. 
This is because we can decompose $N$ into a collection of simple
networks $B_N$ (one for each blob $B$ of $N$) each having at least 2 leaves, and if
each of these simple networks is tree-based, then so is $N$. 
Moreover, for each of these simple networks $B_N$, if we 
remove all but 2 leaves from $B_N$ and obtain a tree-based network, then it is 
straight-forward to see using Lemma~\ref{lem:andrew}
that $B_N$ must have been tree-based.

\begin{figure}[h]
	\begin{center}
		\includegraphics[width=10cm]{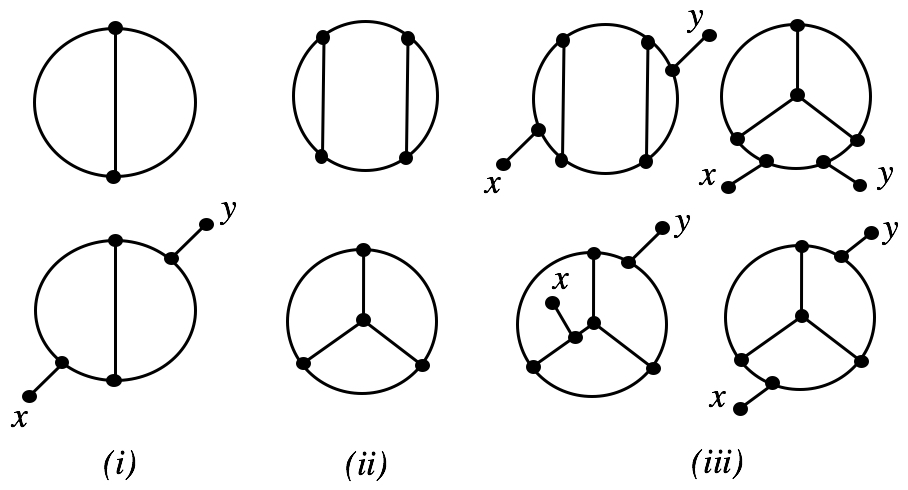}
	\end{center}
	\caption{Diagram for proving that simple level-2 and level-3 networks are tree-based used in the proof of Theorem~\ref{generators}.}\label{generators}
\end{figure}

Now, to see that any simple, level-4 network with two leaves $x$ and $y$
is tree-based, we begin with the case $k=2$.
It is clear that any simple level-2 network 
on some 	set $Y$ can be obtained by inserting pendant edges in 
the multigraph at the top of Figure~\ref{generators}(i) and 
labelling the leaves by the elements of $Y$  (see e.g. \cite[Figure 4]{lm17}).
It is now straight-forward to check that, up to isomorphism, 
the only possible simple level-2 network on $\{x,y\}$ is isomorphic 
to the one at the bottom of Figure~\ref{generators}(i). Clearly, this 
network is tree-based.

We now consider the case $k=3$. 
As before, it is known that any simple level-3 network on some finite set $Y$ can be obtained by inserting
a pendant edge into one of the multigraphs in 
Figure~\ref{generators}(ii) and labelling the leaves  
by the elements of $Y$ (\cite[Figure 4]{lm17}). It is now 
straight-forward to check that the only possible 
simple level-3 networks on $\{x,y\}$ are isomorphic to one 
of the networks in Figure~\ref{generators}(iii), and that each of
these is tree-based.

\begin{figure}[h]
	\begin{center}
		\includegraphics[width=10cm]{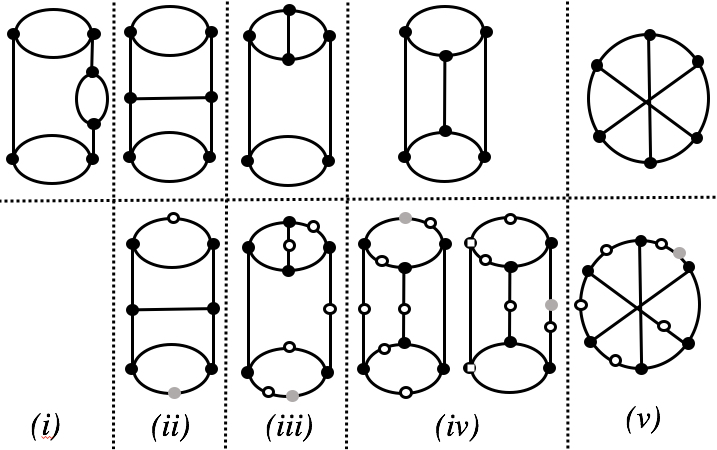}
	\caption{Diagram for proving that simple level-4 networks are tree-based used in the proof of 
Theorem~\ref{level-4-tree-based}\label{fig:new-fig}. In the bottom row, each 
grey vertex corresponds to inserting a pendant edge labelled by $x$, and each 
circle vertex corresponds to inserting a pendant edge labelled by $y$, so that a 
network on $\{x,y\}$ is produced (so, for example, there are 5 possible 
networks associated to the diagram in the bottom row of column (iii)).}
\end{center}
\end{figure}

We conclude with the case $k=4$. We use the fact that any simple level-4 
network with leaf-set $Y$ can be obtained by inserting pendant edges in one of the 
five multigraphs in the top row of Figure~\ref{fig:new-fig} and labelling the leaves by the 
elements of Y (see e.g. \cite[Figure 4]{lm17}). Using this fact, it is now straight-forward 
to check that, up to isomorphism, any simple level-4 network
on $\{x, y\}$ is isomorphic to one of the 
networks on $\{x, y\}$ that can be generated from
the bottom row of Figure~\ref{fig:new-fig}
as described in the figure's caption (note that we can exclude 
the case (i) in the bottom row as it is not possible for this to be 
made into a simple network with leaf-set $\{x, y\}$). 
It is now straight-forward to check that each of these networks on $\{x, y\}$ is tree-based,
which concludes the case $k=4$.

\begin{figure}[h]
	\begin{center}
		\includegraphics[width=10cm]{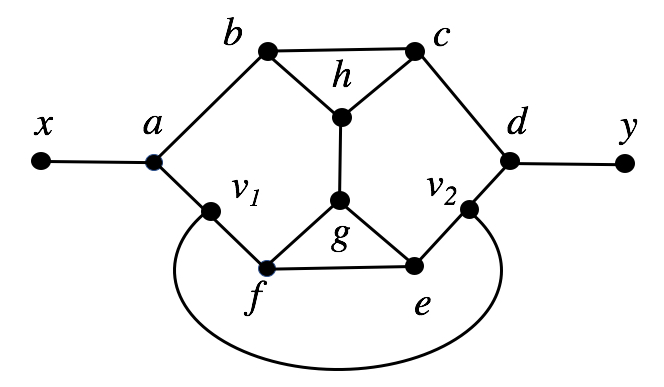}
		\caption{A level-5 network on $\{x,y\}$ 
that is not tree-based \cite{s17}. The labels of the interior vertices
are included for proof purposes.
}\label{f:level5}
	\end{center}
\end{figure}

We now prove the last statement of the theorem.
An example of a level-5 network is presented in Figure~\ref{f:level5}. 
This network can be seen to be not tree-based as follows.  

If it were tree-based, then there would be a path from $x$ to $y$ 
passing through every vertex exactly once.  If the path began $x,a,b$, or 
ended $c,d,y$, then it could not pass through $v_1$ or $v_2$ without 
going through some vertex twice.  Therefore such a path begins $x,a,v_1$ 
and ends $v_2,d,y$.  The edge $\{v_1,v_2\}$ cannot be included in such a path, because 
that completes the path without passing through all vertices, so the path actually 
must begin $x,a,v_1,f$ and end $e,v_2,d,y$.  At this point the path cannot include 
vertices $b,c,h$ without passing through $g$ twice: a contradiction.

This example demonstrates that not all level-5 networks are tree-based, and since by definition of level, this network is also level-$k$ for any $k\ge 5$, the result follows.
\epf

\begin{remark}\label{rem:lev.kge5}
The level-$5$ example used in the proof of Theorem~\ref{level-4-tree-based} can be used to show that it is possible to construct networks that are \emph{strictly} level-$k$ (in that they are level-$k$ and not level-$(k-1)$), and that are not tree-based.  Take a network $N$ of level-$k\ge 5$ and choose a pendant edge 
$\{x,y\}$ in $N$.  Replace this edge by the level-$5$ network shown in Figure~\ref{f:level5}.  The resulting network has unchanged level, and is not tree-based. 
\end{remark}

\section{Recognizing tree-based networks}
\label{sec:recognize}

In this section, we consider the complexity of the 
computational problem of deciding whether
or not a given network $N$ is tree-based.

\begin{figure}[ht]
	\begin{center}
		\includegraphics[width=8cm]{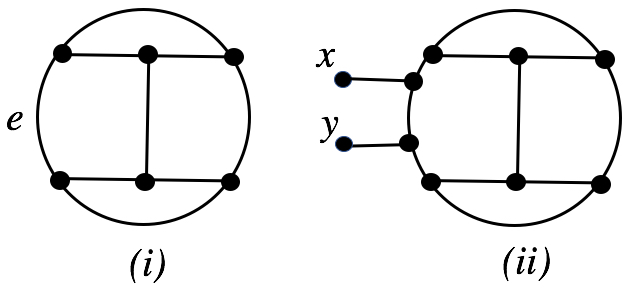}
	\end{center}
	\caption{(i) Cubic graph $C$ with edge $e$ indicated. (ii) The network $C_e(x,y)$.\label{build}}
\end{figure}

We begin with a useful result. Suppose that $C$ is a cubic graph.
Pick some edge $e$ in $C$. Introduce
two pendant edges into $e$ containing the new degree 1 
vertices $x$ and $y$. This new graph $C_e(x,y)$ is a network on
$\{x,y\}$. We illustrate this construction in Figure~\ref{build}.  The following observation 
concerning this construction is straight-forward to check.

\begin{lemma}\label{pair}
Suppose that $C$ is a cubic graph. The following statements are equivalent:\\
(i) $C$ is Hamiltonian.\\ 
(ii) There is some edge $e$ in $C$ such that the network $C_e(x,y)$ is tree-based.\\
(iii) There is some edge $e$ in $C$ such that the network $C_e(x,y)$ has a 
support-tree consisting of a path with end vertices $x$ and $y$.
\end{lemma}

\begin{figure}[ht]
	\begin{center}
		\includegraphics[width=6cm]{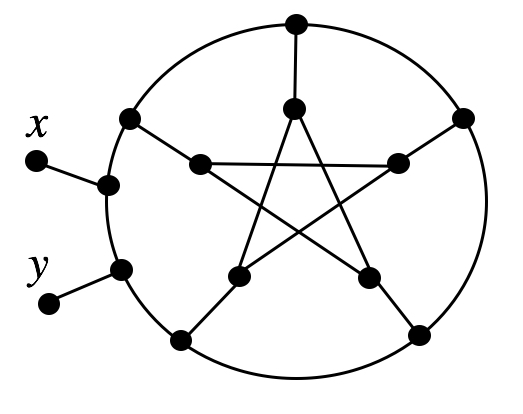}
	\end{center}
	\caption{A simple level-6 network that is not tree-based.
Removing the two pendant edges labelled with $x$ and $y$
and their vertices results in the Petersen graph.}\label{level6}
\end{figure}

Note that using Lemma~\ref{pair}, it immediately follows that the 
network in Figure~\ref{level6} 
is not tree-based, since if $P$ is the Petersen graph (which is not Hamiltonian), then 
this network is of the form $P_e(x,y)$ 
for some edge $e$ of $P$.

We now use the last lemma to prove two NP-completeness results. In our proofs, we shall 
use the fact that the following problem is NP-complete  \cite{g76}:\\

\noindent {\em PC3C-Hamiltonian}\\
\noindent Instance: Planar, cubic, 3-connected graph $G$. \\
\noindent Question: Is $G$ Hamiltonian?\\

We begin by showing that the following problem is NP-complete.\\

\noindent {\em Unrooted tree-based}\\
\noindent Instance: Network $N$ on $X$.\\
\noindent Question: Is $N$ tree-based?\\

\begin{theorem}
	The problem {\em Unrooted tree-based} is NP-complete.
\end{theorem}
\noindent{\em Proof:}
First note that {\em Unrooted tree-based} is in NP since we can check in 
polynomial time if a given tree in $N$ is a spanning tree for $N$ with leaf-set $X$.

To complete the proof, we show that there is a polynomial time reduction from 
{\em PC3C-Hamiltonian} to {\em Unrooted tree-based}.
 
Let $C$ be a  planar, cubic, 3-connected graph. 
By Lemma~\ref{pair}(ii) it follows that  
$C$ is Hamiltonian if and only if the network 
$C_e(x,y)$ on the set $\{x,y\}$ is tree-based for some edge $e$ in $C$.
Since the number of edges in $C$ is equal to $3|V(C)|/2$, 
it follows that there is a polynomial time reduction from 
{\em PC3C-Hamiltonian} to {\em Unrooted tree-based}
(just check if $C_e(x,y)$ is tree-based for each edge $e$ in $C$; if the answer is no
for every $e$ then $C$ is not Hamiltonian, otherwise $C$ is Hamiltonian). 
\epf

Interestingly, the analogous decision problem to 
{\em Unrooted tree-based} for rooted phylogenetic networks 
can be decided in polynomial time \cite{fs15}.

We now  prove that a related decision problem is NP-complete. 
We say that a phylogenetic tree $T$ on $X$ 
is {\em displayed} by a network $N$ on $X$ if 
$T$ can be obtained from a subtree $T'$ of $N$ by suppressing all 
degree 2 vertices in $T'$  \cite{l16}. In addition, we say 
that {\em $T$ is a base-tree 
of $N$} or {\em $N$ is based on $T$} 
if $T$ can be obtained in this way from a support tree $T'$ of $N$.
Note that a phylogenetic tree may be displayed
by a network but need not be a support-tree for the network. 
For example, the phylogenetic tree consisting of an 
edge with leaves $x,y$ is displayed by the 
network in Figure~\ref{build} (e.g consider the 
path of length 3 in the network 
between $x$ and $y$) but it is not a support-tree
for that network. 

We now consider the following decision problem.\\

\noindent {\em Unrooted base-tree containment}\\
\noindent Instance: Network $N$ on $X$ and a phylogenetic tree $T$ on $X$.\\
\noindent Question: Is $N$ based on $T$?\\

Note that the analogous version of this decision problem for rooted 
phylogenetic networks is NP-complete  \cite{A16}. We now 
show that this is also the case for networks. \\

\begin{theorem}
	The problem {\em Unrooted base-tree containment} is NP-complete.
\end{theorem}
\pf
First note that {\em Unrooted base-tree containment} is in NP since we can check in 
polynomial time if a subtree $T'$ of $N$ is a support tree of $N$, and 
that a given phylogenetic tree $T$ can be obtained from
$T'$ by suppressing all degree 2 vertices in $T'$.

To complete the proof, we show that there is a polynomial time reduction from 
{\em PC3C-Hamiltonian} to {\em Unrooted base-tree containment}.

Let $C$ be a  planar, cubic, 3-connected graph.   
By Lemma~\ref{pair}(iii) it follows
that $C$ is Hamiltonian if and only if 
there is some edge $e$ in $C$ such that the network $C_e(x,y)$ has a 
support-tree consisting of a path with end vertices $x$ and $y$ which, in turn,  
holds if and only if the phylogenetic tree $T$ consisting
of a single edge and leaf-set $X=\{x,y\}$ is a base-tree for the network $C_e(x,y)$ 
for some edge $e$ in $C$. Since the number of edges in $C$ 
is equal to $3|V(C)|/2$ it follows that there is a polynomial-time reduction from 
{\em PC3C-Hamiltonian} to {\em Unrooted base-tree containment}
(just check if the single-edged phylogenetic tree $T$ on $\{x,y\}$ is a base-tree 
for the network $C_e(x,y)$ for each edge $e$ in $C$; if the answer is no
for every $e$ then $C$ is not Hamiltonian, otherwise $C$ is Hamiltonian). 
\epf

Note that it is also NP-complete to decide whether 
or not a network $N$ displays a 
phylogenetic tree $T$ \cite{l16}.

\section{Universal tree-based networks}\label{sec:universal}

In this section, we shall show that there are networks on $X$ which
can have {\em every} phylogenetic tree on $X$ as a base-tree. To
prove this we will relate networks with rooted phylogenetic networks,
which we now formally define.

A {\em rooted} phylogenetic network $M$ (on $X$) is a directed acyclic
graph with a single root (vertex with indegree 0 and outdegree 2), 
leaf-set $X$ (vertices with indegree 1 and outdegree 0), 
and all vertices except the root having degree 1 or 3.
If $M$ is a tree, then it is called a {\em rooted} phylogenetic tree on $X$. 
A rooted phylogenetic network $M$ is called {\em tree-based} if it 
contains a directed spanning tree (an `arborescence') $T$ such that 
the leaf set of $T$ is $X$. 
In that case, $T$ is called a {\em support tree} for $M$.

In \cite{h16} it is shown that for every $X$ there exists 
a ``universal'' rooted phylogenetic network $M$ on $X$,
that is, $M$ is a tree-based, rooted phylogenetic network and has every possible rooted phylogenetic
tree on $X$ as a base-tree. We shall use this result 
to show that there are also universal networks. First, we present 
a relationship between tree-based networks and rooted phylogenetic  
networks (cf. also \cite[Section 3]{g12} for related results).

Given a network $N$ on $X$ with $|X|\geq 2$, a leaf $x \in X$, and some 
orientation $o$ of the edges of $N$,
we let $N_o^x$ denote the directed graph which results by 
removing $x$ and its pendant edge from $N$ 
with edges oriented according to $o$.

\begin{theorem}\label{rooted}
	Suppose that $N$ is a network on $X$ and
	$x\in X$. Then $N$ is tree-based if and only if 
	there exists some orientation $o$ of the edges of $N$ making 
	$N_o^x$ a tree-based (rooted) network on $X-\{x\}$.
\end{theorem}
\noindent{\em Proof:}
Suppose $o$ is an orientation of the edges of $N$ such that
$N_o^x$ is a tree-based, rooted network on $X-\{x\}$.
Pick some base-tree $T$ in $N_o^x$. Let $v_x$ denote 
the vertex in $N$ that is adjacent  with $x$ and let $v_1\in V(N)$ 
denote one of the two other vertices in $N$ adjacent with $v_x$.
Then $v_1\in V(T)$. Let $T'$ be the tree obtained from $T$ by first adding
$x$ to its leaf set, $v_x$ to its vertex set, and $\{v_x,x\}$ and $\{v_1,v_x\}$
to its edge set and then ignoring the directions of the edges of $T$.
Since $T$ is a spanning tree of $N_o^x$ with leaf set $X-\{x\}$, we clearly
have that $T'$ is a spanning tree for $N$ with leaf set $X$. Thus, $N$ 
is tree-based.

Conversely, suppose that $N$ is tree-based. Pick some support-tree 
$T$ in $N$ and orient all edges in 
$T$ away from $x$. Let $x,v_1,v_2,\dots,v_m$ be some topological 
ordering of the vertices in $T$ (i.\,e.\, an  ordering that is consistent with the partial ordering induced by $T$). For each vertex $v$ in $T$ that is the 
end vertex of some 
edge in $N$ not in $T$, starting with a vertex that comes earliest in 
the ordering, direct the edge away from $v$, and 
if such an edge is
encountered that has already been directed, then ignore this. 
This choice $o$ of orientations of the edges of $N$ implies that $N_o^x$ 
is a rooted phylogenetic network  on $X-\{x\}$ 
(since it has no directed cycles) with support-tree $T_{o'}^x$ where 
$o'$ is the orientation of the edges induced by $o$.  
\epf

\begin{corollary}
	There exists a tree-based 
	network $N$ on $X$ such that 
	every phylogenetic tree $T$ on $X$ is a base-tree for $N$.
\end{corollary}
\noindent{\em Proof:}
The case $|X|=1$ and $|X|=2$ are obvious. Assume $|X|\ge 3$.
Let $x \in X$ and set $Y =X -\{x\}$. Let $M$ be a universal 
rooted network on $Y$ (see \cite{h16} for details). Let 
$\rho$ denote the root of $M$. Let $N$ be the 
network on $X$ obtained by adding $x$ to the leaf set of $M$,
a new vertex $r$ to the vertex set of $N$, new edges $\{r,x\}$
and $\{r,\rho\}$ to $M$, and ignoring the orientations of all
edges of $M$. Clearly, $N$ is a network on $X$ and
$M$ and $N_o^x$ are isomorphic where $o$ is the 
orientation of the edges of $M$. By Theorem~\ref{rooted},
$N$ is tree-based.
Moreover, if $T$ is any rooted phylogenetic tree on $Y$ then
$T$  is a base-tree for $M$ because $M$ is a universal network on $Y$.
Hence, the tree $T_x$ obtained by 
adjoining the element $x$ as a leaf to the root of $T$
is a phylogenetic tree on $X$ and
ignoring its edge orientations renders it 
a base-tree for $N$. But it
is straight-forward to check that the set
$$
\{ T_x \,:\, x\in X \mbox{ and } T \mbox{ a rooted phylogenetic tree on } X-\{x\} \}
$$ 
is equal to the set of phylogenetic trees on $X$. The
corollary follows immediately.
\epf

\section{Fully tree-based networks}\label{sec:fully-tree-based}

In \cite{s16} a characterization of 
rooted phylogenetic networks in which every embedded 
phylogenetic tree with the same 
leaf-set is a base-tree is given (these are precisely 
the ``tree-child'' networks). In our last result, we will characterize 
networks that have an analogous property. 

Note that a network $N$ on $X$ always contains a subtree with the same
leaf-set as $N$. For example, if we fix some $x \in X$ and 
let $p_{xy}$ be some path in $N$ for all $y \in X -\{x\}$, 
then the tree obtained by removing (if necessary) 
a minimum number of edges from the union of the paths
$p_{xy}$ over all $y \in X-\{x\}$ is a subtree of $N$ with leaf-set $X$.

We call a network $N$ on $X$ {\em fully tree-based} if 
every subtree of $N$ with leaf-set $X$ is a support-tree for $N$.
Note that by the previous remark, any fully tree-based network is tree-based.

\begin{lemma}\label{delete}
	Suppose that $N$ is a simple, tree-based network and 
	that $T$ is a base-tree for $N$. 
	If $e=\{v_1,v\},e'=\{v,v_2\}$ are incident edges in $T$
	such that neither $e$ nor $e'$ are pendant edges of $T$, and $T_e$ and $T_{e'}$ are the trees 
	which are obtained by deleting $e$ and $e'$, respectively, that do not contain $v$, then there
	must exist some edge $e''$ in $N$ which has one vertex in $T_e$ and the other in $T_{e'}$.
\end{lemma}
\noindent {\em Proof:}
Suppose that there is no edge $e''$ with the stated properties. 
Then there exists only one path in $N$ between $v_1$ and $v_2$. But 
this contradicts the fact that $N$ is simple, and therefore the graph $N$ with all 
pendant edges  removed is 2-connected.
\epf

\begin{lemma}\label{exists}
	Suppose that $N$ is a simple, level-$k$ network, $k\ge 1$, on $X$, $|X| \ge 2$. 
	Then $N$ contains a vertex which is not contained in a pendant edge of $N$ if and
	only if $k \neq 1$.
\end{lemma}
\noindent {\em Proof:}
Since $N$ is simple, $|V(N)|=2(|X|-1+k)$ (cf. e.g. \cite{hmw16}). Now, let
$q$ be the number of vertices in $N$ which are not contained in any pendant edge of $N$. 
Then clearly
$$
|V(N)|= 2|X|+ q.
$$
Therefore, $q=2k-2$. The lemma now follows immediately.
\epf

We now characterize fully tree-based networks. Note that these
are significantly less complex than the tree-child networks mentioned above.

\begin{theorem}
Suppose that $N$ is a network on $X$. Then $N$ is fully tree-based if and only if $N$ is a level-1 network.
\end{theorem}
\noindent {\em Proof:}
The statement is clearly true if $|X|=1$. So we assume from now on that $|X|\ge 2$.

By Proposition~\ref{decompose}, it suffices to assume that $N$ is simple.

If $N$ is a simple, level-1 network, then it is straight-forward to check that it is fully tree-based.

Conversely, suppose for contradiction that $N$ is a
simple network on $X$ which is not level-1, 
and that $N$ is fully tree-based. Let $T$ be a support-tree for $N$. 

Suppose that $v$ is a vertex in $T$ that is not contained in some pendant edge of $N$.
Note that such a vertex exists by Lemma~\ref{exists} since $N$ is not level-1.

If the degree of $v$ in $T$ is 2, then 
let $e=\{v_1,v\},e'=\{v,v_2\}$ denote its incident edges neither of which
can be a pendant edge in $N$.
Then, by Lemma~\ref{delete}, we can remove edges $e,e'$ from $T$
and add in an edge $e'' \in E(N)$ in between a vertex of 
$T_e$ and  a vertex of $T_{e'}$ where $T_e$ and $T_{e'}$
are as in the proof of that lemma.
Since the degree of $v$ in $T$ is 2, the resulting tree $T'$ has leaf set $X$. 
Moreover $T'$ does not contain the vertex $v$. 
But this contradicts the fact that $N$ is fully tree-based.

If the degree of $v$ in $T$ is 3, then let $e=\{v_1,v\},e'=\{v,v_2\}$
be two edges incident with $v$. Then by Lemma~\ref{delete}, there is an 
edge $e''$ between a vertex of  $T_e$ and a vertex of $T_{e'}$.
Now, if we remove $e$ from $T$ and add in edge $e''$ we obtain a 
new tree $T'$ that is a support-tree for $N$ and which contains a vertex (namely $v$) with
degree 2, such that neither of the edges in $T'$  incident with
$v$ is a pendant edge of $T'$. But this is impossible by the argument presented above. 
\epf

\section{Final Remarks}\label{sec:conclusion}

We have proven various results concerning tree-based networks, and
shown that they have somewhat different properties as compared
with rooted tree-based networks. 

The structure of rooted tree-based networks is very well understood 
see e.g. \cite{fss17,z16}. It would be interesting to know if
related structural results can be proven for tree-based networks.
In addition, results have recently appeared concerning the
structure of non-binary rooted, tree-based networks \cite{ji16}
(networks in which internal vertices do not necessarily have degree 3).
It would therefore also be of interest to consider properties of non-binary 
tree-based networks.\\

\noindent{\bf Acknowledgments.}
We thank Mike Steel for providing the level-5 network which is
not tree-based pictured in Figure~\ref{f:level5}. We also 
thank the anonymous referee for their helpful comments.
KTH and VM thank the London Mathematical Society for its support and also 
the Centre for Research in Mathematics at Western Sydney University, Australia,
where this work was first discussed. All authors 
thank the Royal Society for its support.

\end{document}